\documentclass[12pt,preprint]{aastex}
\usepackage{natbib}
\usepackage{amsmath}
\def\ovi{{{\rm O}\,{\sc vi}~}}                                
\def\ovii{{{\rm O}\,{\sc vii}~}}
\def\oviii{{{\rm O}\,{\sc viii}~}}

\def\neix{{{\rm Ne}\,{\sc ix}~}}
\def\chandra{{\it Chandra}~}

\def\xmm{{\it XMM-Newton}~}
\def\xmmn{{\it XMM-Newton}}                              
\def\suzaku{{\it Suzaku}~}

\def\msun{{M$_{\odot}$}}

\def\mrk509{{\it Mrk 509}~}
\def\kms{{km s$^{-1}$}}

\title{Probing the Anisotropy of the Milky Way Gaseous Halo-II:
  sightline toward Mrk509}

\author{ A.~Gupta\altaffilmark{1,2}, S.~Mathur\altaffilmark{2, 3}, and 
Y.~Krongold\altaffilmark{4}}

\altaffiltext{1}{Department of Biological and Physical Sciences, 
                      Columbus State Community College, 
                      Columbus, OH 43215, USA;
                      agupta1@cscc.edu}

\altaffiltext{2}{Department of Astronomy, 
		The Ohio State University, 
		140 West 18th Avenue, 
		Columbus, OH 43210, USA}

\altaffiltext{3}{Center for Cosmology and Astro-Particle Physics, 
		The Ohio State University, 
		140 West 18th Avenue, 
		Columbus, OH 43210, USA}

\altaffiltext{4}{Instituto de Astronomia,
                 Universidad Nacional Autonoma de Mexico           
                 Mexico City, (Mexico)}

\begin{document}

\begin{abstract}
  Hot, million degree gas appears to pervade the Milky way halo,
  containing a large fraction of the Galactic missing baryons. This
  circumgalactic medium (CGM) is probed effectively in X-rays, both in
  absorption and in emission. The CGM also appears to be anisotropic, so
  we have started a program to determine CGM properties along several
  sightlines by combining absorption and emission measurements. Here we
  present the emission measure close to the \mrk509 sightline using new
  \suzaku and \xmm observations. We also present new analysis and
  modeling of \chandra HETG spectra to constrain the absorption
  parameters. The emission measure in this sightline is high,
  EM$=0.0165\pm0.0008\pm0.0006~$cm$^{-6}~$pc, five times larger than the
  average. The observed \ovii column density N(\ovii)$= 2.35\pm0.4
  \times 10^{16}$cm$^{-2}$, however, is close to the average. We find
  that the temperature of the emitting and absorbing gas is the same:
  $\log T (\rm K) = 6.33\pm0.01$ and $\log T (\rm K)=6.33\pm0.16$
  respectively. We fit the observed column density and emission measure
  with a $\beta-$model density profile. The the central density is
  constrained to be between $n_0=2.8$--$6.0\times 10^{-4}$ cm$^{-3}$ and
  the core radius of the density profile has a lower limit of 40 kpc.
  This shows that the hot gas is mostly in the CGM of the galaxy, not in
  the Galactic disk. Our derived density profile is close to the
  \citet{Maller2004} profile for adiabatic gas in hydrostatic
  equilibrium with an NFW dark matter potential well. Assuming this
  density profile, the minimum mass of the hot CGM is $3.2 \times
  10^{10}~$M$_{\odot}$.
\end{abstract}

\section{Introduction}

We have known for decades that stellar and ISM components of galaxies,
including our own Milky Way, account for only a small fraction of their
baryons, compared to the amount expected in their halos from the
universal baryon fraction $\Omega_{b}/\Omega_{m} = 0.17$
\citep{Sommer2006, Bregman2007, Anderson2010, Gupta2012, Miller2013,
  Miller2016}. As interesting as the missing baryons problem is the
missing metals problem; nearby galaxies are also short of metals
expected from the star formation history of the universe
\citep{Shapley2003}. Perhaps a solution to both of these problems lies
in the highly ionized warm-hot gas in the circumgalactic medium (CGM) of
galaxies. Recent theoretical models suggest that the CGM of galaxies 
contains a large reservoir of warm-hot gas accounting for the majority
of galactic baryons \citep{Stinson2012, Fang2013, Feldmann2013,
  Ford2013, Roca-Fabrega2016} and according to \citet{Peeples2014}, the
CGM could account for 40\% of metals produced by star forming galaxies.
The distribution, spacial extent, and mass of this warm-hot gas provide
important constraints to models of galaxy formation and the accretion
and feedback mechanisms.

Although theoretical models predict the existence of the warm-hot gas in
the CGM, detecting and characterizing the diffuse CGM has been
difficult.  Because of our special vantage point, our own Milky Way
provides a unique opportunity to probe the CGM of a spiral galaxy.  The
warm-hot CGM gas can be effectively probed by highly ionized metals; the
dominant transitions from such ions lie in the soft X-ray band. In
literature there are multiple reports on detection of redshift zero
absorption lines due to \ovii and \oviii 
\citep{Nicastro2002,Wang2005,Williams2005,Fang2006,Bregman2007,
Miller2013,Fang2015}. But it is difficult to measure
the extent, density, and mass of this warm hot gas because of the
inherent difficulty in using absorption line studies alone. While the
absorption line column density is a product of density and pathlength,
emission measure is a product of density square and pathlength, so a
combination of absorption and emission studies are required to break the
degeneracy and so fully characterize the warm-hot CGM. Indeed various 
broad-band X-ray observations have revealed an extensive 
diffuse soft ($\le$ 1 keV ) X-ray background (SXRB). Shadow observations 
show that there is a significant contribution from the Galactic halo to SXRB 
\citep{Snowden2000,Smith2007,Galeazzi2007,Henley2007,Henley2008,Gupta2009}. 

In \citet[hereafter Paper-I]{Gupta2012}, 
combining \chandra observations of \ovii and \oviii
absorption lines and \xmm and \suzaku measurements of the Galactic halo
emission measure, we found that there is a huge reservoir of ionized gas
around the Milky Way, with a mass of over 60 billion solar masses and a
radius of over 100 kpc. Thus there appears to be more baryonic mass in
the warm-hot CGM than in the entire disk of the Galaxy and as much mass
in metals as in all the stars in the disk. In Paper-I we compared
absorption and emission values average over the whole sky. However,
shadow observations and other studies of soft X-ray diffuse background
show that the emission measure of the Galactic halo varies by an order
of magnitude in different sight-lines
\citep{Henley2010,Gupta2009}. Therefore it is crucial to determine the
emission measure of emitting gas near absorption sight-lines to
understand the differences in physical properties of the CGM across the
sky.

In \citet[hereafter Paper-II]{Gupta2014} we compared absorption and
emission along two sightlines: toward Mrk421 and PKS2155-304.  In these
two sightlines, the observed column densities are similar, but observed
emission measures are different, so their densities and/or pathlengths
must be different. Indeed we found that toward Mrk421 and PKS2155-304 
the densities are $1.6_{-0.8}^{+2.6} \times 10^{-4}~$cm$^{-3}$ 
and $3.6_{-1.8}^{+4.5} \times 10^{-4}~$cm$^{-3}$ and pathlengths are 
$334_{-274}^{+685}~$kpc and $109_{-82}^{+200}~$kpc respectively. While the 
errors on the derived parameters are large, this provides a
suggestive evidence that the warm-hot gas in the CGM of the Milky Way is
not distributed uniformly.

Here we expand on our previous work and constrain the physical
properties (temperature, path-length, density, and mass) of CGM in the
sightline toward \mrk509. In sections 2 and 3 we present a detailed
analysis of our \xmm and \suzaku (PI: Gupta) new observations of a blank
sky field near \mrk509. We also present detailed reanalysis of the
\chandra High Energy Transmission Grating (HETG) 2012 observation of
\mrk509. \citet{Kaastra2014} have analyzed and modeled the intrinsic
absorbers of \mrk509. Here we focus on the redshift zero absorption
lines. In section 5 we present results with a uniform density model as
well as a $\beta-$model. The discussion is presented in \S 7.

\section{Observation and data reduction}

A blank X-ray sky field ({\it Off-field2}) adjacent to \mrk509
was observed by \xmm and \suzaku.  The observation IDs, dates, pointing
directions and exposure times are summarized in Table 1. Figure~1 shows
the ROSAT All Sky Survey (RASS; 0.1-2.4 keV) image in the vicinity 
of \mrk509, along with the \xmm and \suzaku pointing of {\it Off-field2}.
\mrk509 was observed with \chandra Low Energy Transmission Grating
(LETG) and High Energy Transmission Grating (HETG) in December 2009 and
September 2012 respectively. We presented our results on the $z=0$ \ovii
and \oviii absorption lines from the LETG observation in Paper-I.
In the following we report on the data reduction of \xmm and
\suzaku observations of {\it Off-field2} and \chandra HETG observation
of Mrk~509.

\subsection{XMM-Newton observations}

{\it Off-field2} was observed by XMM-Newton for 60 ks on November
2013. The observation was performed with the “thin” filter using the 
full-frame (FF) mode for all the EPIC (European Photon Imaging Camera) 
cameras. In this work we used data from the EPIC-pn and EPIC-MOS
detectors. 


We reduced the data using the XMM-Newton Extended Source Analysis 
Software\footnote{\url{ftp://xmm.esac.esa.int/pub/xmm-esas/xmm-esas.pdf}} 
(XMM-ESAS; Snowden \& Kuntz 2013) as distributed with version 13.5.0 of the 
Science Analysis System (SAS). We first used the standard SAS 
{\it emchain} and {\it epchain} scripts
to produce calibrated events lists from MOS and pn cameras respectively, 
and then run the XMM-ESAS {\it mos-filter} and {\it pn-filter} scripts. 
{\bf These scripts call the SAS task {\it espfilt} for light-curve cleaning. 
It uses broad-band ($2.5-8.5~$ keV) count rate to filter flares.} 
This process mostly removed the soft proton contamination, but there 
could be some residual emission left. The resultant good exposure time 
after this filtering are $45.6~$ks, $47.6~$ks, and $34.2~$ks for the MOS1, 
MOS2 and pn cameras, respectively. Note that the resulting pn exposures 
are shorter than the corresponding MOS exposures. 
As reported in ESAS cookbook, this is likely due 
to the greater sensitivity of the pn detector, 
meaning that relatively smaller departures from the mean count rate are 
flagged as soft proton flares.

Since we are interested in the diffuse emission, we need to remove any
contribution to emission from point sources in the field. We detected
the point sources and created the source list using the XMM-ESAS task
{\it cheese}. The {\it cheese} task combines both MOS and pn data for source
detection creating images and exposure maps in a selected single band
($0.5$--$2.0$ keV). We have selected PSF threshold parameter of {\it
  scale=0.2} which corresponds to point source removal down to a level
where the surface brightness of a point source is 20\% of the
surrounding background. We set the detection limit (flux in the energy
band of $0.5$--$2.0~$KeV) of $1 \times 10^{-15}~$erg cm$^{-2}$ s$^{-1}$
to create masks in order to remove the point sources. After
removal of bad pixels and point sources, the active fields of view are
$5.0 \times 10^{-5}~$sr and $4.8\times 10^{-5}~$sr for the MOS2 and pn
respectively.

We used the XMM-ESAS scripts {\it pn-spectra} and {\it mos-spectra} to
extract spectra from the active fields of view. The spectral extraction
scripts also calculated the redistribution matrix file (RMF) and the
ancillary response file (ARF) for each spectrum, using the SAS {\it
  rmfgen} and {\it arfgen} tools, respectively.  We run the scripts {\it
  mos\_back} and {\it pn\_back} to calculate corresponding quiescent
particle background (QPB) spectra.  The QPB spectra were calculated from
a database of filter-wheel-closed data, scaled to our observations using
data from the unexposed regions of the cameras (Kuntz \& Snowden 2008).
Before our spectral analysis, we grouped each spectrum
such that there are at least 50 counts per bin, and subtracted the
corresponding QPB spectrum (Fig. 2).

\subsection{Suzaku observations}

{\it Off-field2} was also observed with \suzaku for 61~ks on May 2012. 
The data reduction and analysis were carried out with HEAsoft version 6.16 
and XSPEC 12.8.2 with AtomDB ver.2.0.2. In this work, we analyzed only 
the X-ray Imaging Spectrometer (XIS1) data, as 
this has the greatest sensitivity at low energies. We combined the data taken 
in the $3 \times 3$ and $5 \times 5$ observation mode using the ftool 
{\it $xis5\times5to3\times3$}. We first converted $5 \times 5$ mode data to 
$3 \times 3$ mode data and then merged both files with the help of 
ftool {\it ftmerge}. Along with standard data processing 
(e.g. ELV $> 5$  and DYE ELV $> 20$), 
we expanded data screening with the cut-off-rigidity (COR) of the Earth's 
magnetic field, which varies as \suzaku traverses its orbit. 
During times with larger COR values, fewer particles are able to 
penetrate the satellite and the XIS detectors. 
We excluded times when the COR was less than 8 GV, which
is greater than the default value (COR 4 GV), as lowest possible background 
is desired. 

Due to {\it Suzaku's} broad point spread function (half-power diameter 
$\sim$2' \citep{Mitsuda2007}), it is hard to detect point sources. 
Therefore we used the location of point sources determined in the XMM-Newton
observation and excluded a region of 2' radius around the source location. 

We run ftool {\it xisrmfgen} and {\it xissimarfgen} to produce the 
RMFs and ARFs. For the ARF calculations we assumed a uniform source of radius 
$20^{\prime\prime}$ and used a detector mask which removed the bad pixel 
regions. We extracted the spectra of non-X-ray background from a database 
of the night Earth data with ftool {\it xisnxbgen}. 

\subsection{Chandra Observations}

\mrk509 was observed by \chandra HETG in 2012 
for a total of $280~ks$ (ObsID: 13864 \& 13865). The HETG is comprised of two
gratings: the medium energy gratings (MEG) and the high energy gratings
(HEG), which disperse spectra into positive and negative spectral
orders. Since we are interested in oxygen absorption lines we 
used the data from MEG only and reduced it using the standard Chandra
Interactive Analysis of Observations (CIAO) software (v4.6) and Chandra
Calibration Database (CALDB, v4.6.3) and followed the standard Chandra
data reduction threads\footnote{\url{http://cxc.harvard.edu/ciao/
threads/index.html}}. First we run the \emph{mkgrmf} CIAO script to create 
the first order positive and negative MEG RMFs, needed for spectral analysis 
of grating observations. Further, to increase the S/N we co-added the negative 
and positive first-order spectra with \emph{add\_grating\_orders} and built 
the ARFs using the \emph{fullgarf} CIAO script.

\section{Spectral Analysis: Emission} 

We used Xspec for spectral analysis of \xmm and \suzaku data.  We used
the $\chi^{2}$ statistics as a goodness-of-fit measure and all errors
are given at the $1\sigma$ confidence level. The goal of our \xmm and
\suzaku observations of {\it Off-field2} is to measure the contribution
of galactic halo emission to the soft diffuse X-ray background (SDXB)
near \mrk509. SDXB spectra have three distinct components: (1) a
foreground component consisting of solar wind charge exchange (SWCX) and
the local bubble (LB); this is modeled as an unabsorbed plasma with
thermal emission in collisional ionization equilibrium (CIE); (2) a
background component made of unresolved extragalactic sources; this is
modeled with an absorbed power law; and (3) Galactic halo emission; this
is modeled as an equilibrium thermal plasma component absorbed by the
gas in the Galactic disk. 

Both the Galactic halo and foreground components (SWCX+LB) have similar
spectral shape, primarily X-ray lines from highly ionized metals, mostly
\ovii and \oviii. As a result, disentangling the two is very difficult.
To minimize the SWCX contribution, which varies both in spectral
composition and flux on scales of hours to days, we use proton flux
filtering \citep{Smith2007, Yoshino2009}. We obtained the solar wind
proton flux data from OMNIWeb\footnote{http://omniweb.gsfc.nasa.gov/}.
Figure 3 shows the solar wind proton flux during \xmm and \suzaku
observations of {\it Off-field2}. During the \suzaku observation, solar
wind proton flux is much higher than during the \xmm observation. We
investigated the effect of the higher proton flux during \suzaku
observation by comparing surface brightness of \ovii K$\alpha$ and
\oviii K$\alpha$ lines in \xmm and \suzaku spectra. As expected, \ovii
K$\alpha$ line intensity is higher by about a factor of 1.5 during the
\suzaku observation compared to the \xmm observation (Table 2), implying
that the \suzaku spectra have significant contamination from SWCX. Thus
for further analysis we use only the \xmm spectrum of {\it Off-field2}.

\subsection{Spectral Modeling: Emission}

We {\bf simultaneously} fit the \xmm MOS1 and MOS2 filtered diffuse background 
spectra with the three component model noted above. 
Even after the data cleaning described in \S 2.1, there may remain 
some residual soft proton contamination (Snowden \& Kuntz 2011). This 
is modeled as an additional powerlaw that was not folded through the 
instrumental response. We also modeled the instrumental 
Al and Si fluorescence lines at 1.49 and 1.74 keV respectively, with two 
gaussians. 

We use the {\it APEC} model with plasma temperature of 
$T = 1.2 \times 10^{6}~$K
(frozen) and emission measure (EM) of 0.0032~cm$^{-6}~$pc (frozen)
for the foreground component (SWCX plus LB). We determined the
normalization/EM of the foreground component using data from 
\citet{Snowden2000} catalog of SXRB shadows as described in detail in 
Paper-II. We find 5 shadows in the catalog closest to our sightline
with average foreground R12 count-rate of $337 \times
10^{-6}~$counts~s$^{-1}~$arcmin$^{-2}$ corresponding to EM of
$0.0032$~cm$^{-6}~$pc .

The contribution from unresolved extragalactic sources is modeled with
an absorbed power-law. The Galactic column density was fixed to $N_{H}= 4
\times 10^{20}~$cm$^{-2}$ \citep{Dickey1990}, and power-law slope
and normalization were left as free parameters in the spectral fit.

Finally we determine the hotter Galactic halo contribution, modeled as
an equilibrium thermal plasma component absorbed by the gas in the
galactic disk (Fig. 4). We measured the galactic halo temperature of 
$\log T (\rm K) = 6.33\pm0.01$ and emission measure of
$0.0165\pm0.0008\pm0.0006~$cm$^{-6}~$pc. The first error indicates the
statistical error, and the second error indicates the estimated
systematic error due to our assumed foreground spectra (for details see 
Paper-II). The EM towards \mrk509 is unusually high, about five times the 
sky average of $0.0030\pm0.0006~$cm$^{-6}~$pc. \citet{Henley2007} have reported 
such high Galactic halo EM towards a filament in the southern 
Galactic hemisphere. 

\xmm pn spectrum is consistent with the MOS model
($\chi^{2}_{\mu}=1.2$), though there are some residuals at energies
$0.4$--$0.7~$ keV.  This might be due to under- or over-estimation of
the pn particle background as discussed at Snowden \& Kuntz (2013). Thus
for further calculations we use only the MOS2 spectrum fit results
(\citet{Henley2010} have also used only the MOS data for galactic halo
emission measurements).

\section{Spectral Analysis: Absorption}

For the spectral analysis of absorption along the sightline to Mrk 509,
we binned the \chandra MEG spectra to 0.01\AA\ and analyzed using the
CIAO fitting package {\it Sherpa.}

Since we are interested in $z=0$ highly ionized metal lines we fit the 
continuum spectrum in the $15 - 23~$ \AA~ range with a powerlaw, 
absorbed by the Galactic column density. \mrk509 is known to have 
intrinsic absorption (Kaastra et al. 2014 and references there in), 
so we fit the intrinsic absorbers of \mrk509 with multiple gaussians.

After fitting the continuum and intrinsic features, the $z=0$ \ovii K$\alpha$, 
\ovii K$\beta$ and \oviii K$\alpha$ absorption lines are detected with 
more than 3$\sigma$ significance (Fig. 5). We fit these lines with 
narrow Gaussian features of line width 1 m\AA. The best-fit line parameters 
and errors (calculated using the projection command in {\it Sherpa}) 
are given in Table 3 and the spectrum is shown in figure 5. 
The measured equivalent widths (EWs) are consistent within 
1$\sigma$ error of our previous 
measurements (Paper-I) done with the 
\chandra LETG 2009 observation. 

The \ovii K$\alpha$ absorption line is clearly saturated in the \mrk509
data. The measured $\frac{EW(K\beta)}{EW(K\alpha)}$ ratio is 0.56, much
higher than the expected ratio of 0.156 for optically thin \ovii
lines. For the saturated lines like we observe here, converting the
observed EWs to column densities is non-trivial. We used the method
described in detail at Paper-I to obtain constraints on the
\ovii column density: $\log N_{OVII}~(cm^{-2})=16.6\pm0.4$ and the velocity
dispersion parameter $b=70$--$200$ \kms. The measurement uncertainties 
are large due to the weak \ovii K$\beta$ line, but we can do better by
using the code PHASE, as discussed below.

\subsection{Fitting with PHASE}

Our hybrid photo- and collisional-ionization code PHASE \citep{Krongold2003} 
automatically takes into account line saturation by using
Voigt profiles to fit absorption lines. Additional advantage of using
the code is that it fits the entire 11.0 to 23 \AA~ spectrum, taking into
account lines which are not individually detected, providing better
constraints. The fit provides constraints on the column density,
velocity dispersion parameter and temperature of the absorbing
plasma. Best fit PHASE parameters are reported in Table 4 and the
spectrum is shown in figure 6. The \ovii column density was found to be
$\log N_{OVII}~(cm^{-2})=16.37^{+0.07}_{-0.08}$, somewhat lower
than, but consistent with that noted above, and with much smaller
errors.

In the Mrk 509 spectrum we could determine the $z=0$ \neix column
density accurately ($\log N_{NeIX}~(cm^{-2})=15.85\pm 0.2$) even
though the line was not individually detected.

The velocity dispersion parameter obtained with PHASE fitting is
$b=74^{+80}_{-74}$ \kms (where the lower limit is pegged at the hard limit)
and temperature $\log (T/k)=6.33\pm0.16$. The $b$-parameter is not well
constrained, but is consistent with the value obtained from using the
K$\alpha$ and K$\beta$ line ratio noted above. We could not constrain the
metallicity of the gas independently in PHASE; assuming
$Z=0.3Z_{\odot}$, the fit gives total equivalent Hydrogen column density
of $N_H= 2.15\times 10^{20}$ cm$^{-2}$ (Table 4).

\section{Results}

While the strength of an absorption line depends on the
ionic column density of the intervening gas ($N_H= \mu n_e R$, where
$n_{e}$ is the electron density and $R$ is the path-length, and $\mu$ is
the mean molecular weight), the emission is sensitive to the square of
the number density ($EM= n_e^2 R$, assuming a constant density
plasma). Therefore a combination of absorption and emission measurements
breaks the degeneracy and provides constraints on physical properties
such as path-length and density of the absorbing/emitting gas. 

From the PHASE model of z=0 absorption lines and galactic halo emission
model, we constrained the temperature of the absorbing and emitting gas
to $\log ({\rm T (K)})=6.33\pm0.16$ and $\log ({\rm T (K)})=6.33\pm0.01$
respectively.  Since the temperature of the absorbing and emitting gas
is the same, it is reasonable to assume that both absorption and
emission arise in the same plasma.  We can now combine the column
density and emission measure to extract physical properties of the
absorbing-emitting warm-hot gas.

\subsection{Uniform Density Halo Model}

Assuming that absorbing/emitting plasma has a constant density we derive 
the density of: 

\begin{equation}
n_e= (6.6^{+1.7}_{-1.2}) \times 10^{-4} (\frac{0.5}{f_{O VII}})^{-1} cm^{-3}
\end{equation}

and the path length of:

\begin{equation}
R = (126^{+43}_{-41}) (\frac{8.51 \times 10^{-4}}{(A_O/A_H)}) 
(\frac{0.5}{f_{O VII}})^2 (\frac{0.3Z_{\odot}}{Z}) ~kpc
\end{equation}

where the Solar Oxygen abundance of $A_O/A_H= 8.51 \times 10^{-4}$ is 
from \citet{Anders1989} and $f_{O VII}$ is the ionization fraction of \ovii. 
Simulations of the CGM around disk galaxies \citep{Sommer2006,Toft2002} 
suggest the mean metallicity of gas is 
$Z=0.2 \pm 0.1 Z_{\odot}$. These values of metallicities are also 
consistent with observational results for the outskirts of groups 
\citep{Rasmussen2009}, clusters of galaxies \citep{Tamura2004} 
and external galaxies \citep{Anderson2016,Bogdan2013a}. 
As path length is inversely proportional to the  metallicity (equation 2), 
lower values of  metallicity corresponds to larger path-length and 
consequently the higher mass. Thus to be conservative, and 
consistent with our Paper-I and Paper-II, we used $Z=0.3 ~Z_{\odot}$.

The CGM parameters along just one sightline are presented here, and as
noted in \S 1 the Milky Way CGM is likely anisotropic. So we cannot
determine the CGM mass using parameters of only one
sightline. Nonetheless, for the sake of comparison with another density
distribution such as the $\beta$-model discussed below, we will assume
spherical symmetry and covering fraction of 1 of the warm-hot CGM gas
with density and pathlength calculated above. This leads to the mass of
the plasma $=5.4\times 10^{10}~$\msun. For a covering factor of $0.72$
used in Paper-I, the mass is $=3.9\times 10^{10}~$\msun. This is lower
then the mass derived in Paper-I ($=2.3\times 10^{11}~$\msun); this is
largely due to the unusually high EM along this sightline, which leads
to larger density and smaller pathlength (\S 7).

\subsection{$\beta$-model}

Above (and also in Paper-I and Paper-II) we had assumed a constant
density plasma, but the Galactic halo likely has a non-uniform density
profile.  Since the EM is biased toward high density, this could affect
the results. A realistic halo is likely to have some density profile
that falls with radius, so we should use a model with a reasonable density
distribution and compare the model predictions to observations for
obtaining constraints on the structure of the Galactic CGM.

The distribution of the warm-hot gas around several elliptical galaxies
follows a beta-model profile in which the density is high in the center
and falls off with radius \citep{Forbes2012, Mathur2008}; gas in
clusters and groups of galaxies also follows similar profile
\citep{Jones1984, Forman1985, Mulchaey1998}. \citet{Anderson2011}, 
\citet{Anderson2016} and \citet{Dai2012} used the $\beta$ model to fit the
  radial surface brightness profile of hot halo gas around spiral
  galaxies.

The $\beta$-model is given by 
 
$$
n(r) = n_0 [1+ (r/r_c)^2]^{-3\beta/2}
$$

where $r$ is the galactocentric radius, $n_{o}$ is the central density,
$r_{c}$ is the core radius and $\beta$ describes the shape.  We have two
measurements (column density and emission measure) and three unknowns
($n_{o}$, $r_{c}$ and $\beta$), so the parameters are degenerate. We
first assumed $\beta$=0.5 in our calculations and then varied the value
of $\beta$ to see its effect on the results.

Our sightline to \mrk509 passes through the Milky Way halo along the direction
fixed by the Galactic coordinates ($l=35.97^{\circ}$,
$b=-29.85^{\circ}$), and our measurements of absorption and
emission are along this sightline. So we first converted $n(r)$ to
$n(s)$ where $s$ is the pathlength along the sightline. We assumed that
we are at 8~kpc from the Galactic center and the virial radius of the
Galaxy is $250~$kpc, providing the maximum pathlength for
integration. Using {\it Mathematica}, we then determined the values of
$n_{o}$ and $r_{c}$ which are consistent with both the absorption column
density ($N_H$) and the emission measure ($EM$).

The results are shown in figure 7, where $n_{o}/\mu $ is plotted on the
X-axis and $r_{c}$ on the Y-axis. Solid lines mark the best-fit and
error contours for the observed column density and the dash lines are
the same for the emission measure. The central density is constrained to
be between $n_{o}=2.8$--$6.0 \times 10^{-4}$ cm$^{-3}$. For the core
radius, however, we only have a lower limit $r_{c} \geq 40$ kpc. For
larger values of $\beta$, the lower limit on $r_{c}$ increases ($r_{c}
\geq 50$ kpc for $\beta=0.6$), the upper limit on $n_{o}$ becomes
slightly smaller ($n_{o}=5.8 \times 10^{-4}$ cm$^{-3}$ for $\beta=0.6$),
but the lower limit on $n_{o}$ remains the same.

For $r_{c}=40~$kpc (lower limit) and the corresponding $n_{o}=0.0006$
cm$^{-3}$, the mass within the virial radius of 250 kpc is $3.2\times
10^{10}$\msun. For the lower limit of density ($n_{o}=0.00028$
cm$^{-3}$) and the corresponding $r_{c}=280~$kpc, the mass is
$9.98\times 10^{10}$\msun (Table 5).

\section{Comparison with External Galaxies}

Similar to the Milky Way, other spiral galaxies should also have
massive, extended reservoirs of ionized hot gas in the CGM. X-ray
absorption line spectroscopy of galactic haloes is difficult because
only a small number of AGNs are bright enough, but several authors have
studied the halos around spiral galaxies in emission. Most of these,
however, are massive spiral galaxies, not Milky way-type.

\citet{Anderson2016} and \citet{Dai2012} detected hot gaseous halos in
emission around the giant spiral galaxies NGC~1961 and UGC~12591,
extending to 40-50 and 110 kpc respectively. These authors estimates the
mass of their hot halo gas to be $5 \times 10^{9} M_{\odot}$ and $3.9
\times 10^{9} M_{\odot}$ respectively within a radius of 50 kpc, and
when density profiles are extrapolated to virial radii of 500 kpc, the
implied hot halo mass is $1-3 \times 10^{11} M_{\odot}$.  Similarly
\citet{Bogdan2013a, Bogdan2013b} have detected X-ray emission around two
other spiral galaxies NGC~6753 and NGC~266 and estimated the hot X-ray
gas mass within $\sim60~$kpc to be $1.2 \times 10^{10} M_{\odot}$ and
$9.1 \times 10^{9} M_{\odot}$, respectively.  Though the detected masses
in hot halos of these galaxies are significant, it is not a major
contributor to the galactic missing baryons because these are very
massive galaxies; it falls short by an order of magnitude.

Around low mass (Milky Way type) spiral galaxies, the picture is a bit 
different.  \citet{Strickland2004} 
using \chandra observations found diffuse X-ray emitting halos in 
eight nearby ($D<17~$Mpc) galaxies extending to radii of $18~$kpc.
Since these are very nearby galaxies, the \chandra field of view probes 
only a 20~kpc region around the galaxies. With \suzaku observation, 
\citet{Yamasaki2009} confirmed the X-ray
halo of NGC~4631 ($D\sim8~Mpc$) extending out to about 10 kpc from 
the galactic disk. With \xmmn, \citet{Tullmann2006} detect the diffuse 
gaseous X-ray halos extended over a range of 4-10 kpc 
around nine nearby star-forming edge-on spiral galaxies. But these 
observations ($8-55~$ks) are not deep enough to detect soft X-ray halos 
extending out to large radii.

Recently \citet{Bogdan2015} searched for the hot coronae around lower
mass spiral galaxies with stellar masses of ($0.7$ -- $2.0$) $\times
10^{11}~$M$_{\odot}$ using \chandra ACIS observations.  They did not
detect a statistically significant hot corona around any of their sample
galaxies. The authors noted that the low effective area of \chandra
ACIS-I or smaller field-of-view of ACIS-S might be reasons for the
non-detections.

So we see that the observations of Milky Way-type galaxies were either
too shallow or with too small a field of view to detect extended,
massive CGMs. Extended, massive CGMs are detected around massive spiral
galaxies, but they do not contribute significantly to the baryon budget.
Why some spiral galaxies have a large fraction of baryons in their CGM
and some do not? The extent of the CGM in a galaxy and the fraction of
missing baryons it contains may depend on several properties of a
galaxy: stellar mass \citep{Bogdan2015}, specific star formation rate
\citep{Tumlinson2011}, or dark matter halo mass
\citep{Oppenheimer2016}. All the three reasons could be related as
galaxies with higher mass have substantially lower specific star
formation rates \citep{Genel2014, Damen2009} and feedback may affect
both the star formation rate and stellar mass.  Radio-mode feedback may
expel gas from a galaxy leading to both smaller stellar mass and less
massive CGM \citep{Bogdan2015}. In the simulations of 
\citet{Roca-Fabrega2016}, the fraction of hot gas mass in more 
massive galaxies is indeed smaller.  

\section{Discussion}

In Paper I we had assumed a constant density model to determine the CGM
parameters of the Milky Way. We found a huge reservoir of baryonic mass
of over $6\times 10^{10}~$\msun. This result was criticized by some who
argued that the mass is overestimated because of the assumption of
constant density \citep[also see Mathur 2012 and references
therein]{Wang2012}. In Paper II we made logical arguments and showed
mathematically that that is not the case. On the contrary, any
non-uniform density profile with density falling with radius would
necessarily lead to larger mass estimates. Here we confirm our arguments
by comparing the constant density model with a $\beta$-model.  Even the
lower limit on mass is comparable to that with a constant density (see
Table 5), and is over twice as much for the lower limit on the central
density.

CGMs also likely have a radial temperature profile, but we do not
  have observational constraints to determine the temperature profile,
  so we assume constant temperature. With more sightlines we will have
  more observables to determine the temperature profile. We note that
  \citet{Faerman2016} also assume that the mean gas temperature is
  constant as a function of radius in their theoretical model. 

  Another matter of contention in literature is the location of the
  warm-hot gas, whether in the ISM of the Galactic disk, the CGM, or
  beyond. In Paper-II we had argued that the Galactic ISM cannot be the
  major contributor to the $z=0$ absorption lines. Here we constrain the
  core radius of the absorbing/emitting gas to be over $40~$kpc, clearly
  ruling out the ISM origin and placing the gas in the CGM of the Galaxy
  (see also Nicastro et al. 2016a).

In Paper-I we had used sky-average values of column density and emission
measure, with $\log N_{OVII}$ (cm$^{-2}$) $=16.19^{+0.08}_{-0.08}$, and
$EM=0.003~$cm$^{-6}~$pc. What we find here, along the \mrk509 sightline
is $\log N_{OVII}$ (cm$^{-2}$) $=16.37^{+0.07}_{-0.08}$, and
$EM=0.0165\pm0.0008\pm0.0006~$cm$^{-6}~$pc. Thus, along the \mrk509
sightline the \ovii column density is larger by 0.2~dex and the emission
measure is larger by a factor of $5$. As a result, the density along
this sightline is larger than average and the pathlength smaller than
average (for a constant density model as in Paper-I). The parameters
along the two sightlines presented in Paper-II are: $\log N_{OVII}$
(cm$^{-2}$) $=16.22\pm0.23$ \& $16.09\pm0.19$, and
$EM=0.0025\pm0.0003\pm0.0005~$cm$^{-6}~$pc \&
$0.0042\pm0.0003\pm0.0007~$cm$^{-6}~$pc, respectively. This underscores
the value of measuring absorption and emission along the same sightline
to derive the physical properties of the warm-hot gas. We should do so
for several sightlines through the MW-CGM to understand its average
properties and anisotropy. Part of the anisotropy would arise from our
location in the Galaxy as different sightlines would probe different
parts of the Galactic CGM \citep[and references
therein]{Nicastro2016b}. But part of the anisotropy could also be
intrinsic, as suggested by simulations of \citet{Roca-Fabrega2016}.
We will investigate this further with more sightlines with absorption
and emission observations.

In figure 8 we have plotted the $\beta-$model profile in the \mrk509
direction together with the profile obtained by \citet{Fang2013}, which
is the \citet{Maller2004} profile. The Maller \& Bullock profile
specifies the density and temperature profiles for adiabatic hot gas
with polytropic index $n=5/3$ in hydrostatic equilibrium in an NFW dark
matter potential well. We see that shape of their profile is similar to
our $\beta-$model profile for the lower limit of $r_{c}$, although our
density is higher. Our lower limit on the central density is similar to
theirs, but our corresponding profile is flatter. Given that the
$\beta-$model we present here is deduced from only one sightline, and
that too with unusually high emission measure, the differences are not
surprising.


Recently \citet{Faerman2016} presented an analytic
phenomenological model for warm-hot CGMs of L$^{\star}$ galaxies. The
hot gas is in hydrostatic equilibrium in a Milky Way gravitational
potential. They find the median temperature of the hot gas to be
$1.8\times 10^6$ K, similar to what we find for the MW CGM. They also
find the CGM to be extended, slightly beyond the virial radius, and
massive with $1.35\times10^{11}~M_{\odot}$, accounting for missing
baryons in galaxies in the local universe. These results, as well as
those from \citet{Fang2013} and other theoretical models noted in \S
1, provide strong support for our results on the Milky Way CGM. 

In these works the density distribution of the warm-hot gas is presented
as a smooth profile. It is possible, however, that the gas is clumpy and
this could have a noticeable impact on model fits because emission
measures scale with $n^2$ while absorption scales with $n$. Recent
theoretical simulations by \citet{Roca-Fabrega2016} show the presence
of filamentary structure in the CGM. In future work when we have
emission and absorption measurements along several directions, we will
include clumping factors $\langle n^2 \rangle/\langle n\rangle^2$ in our
models to assess their impact on our interpretation, and we will look at
hydrodynamic simulations of galaxy formation for guidance about expected
levels of clumping and anisotropy. We note, however, that in contrast to
the cool CGM component traced by high-velocity HI clouds, we expect the
hot gas component detectable in X-ray data to be relatively smooth
\citep[e.g.,][]{Stocke2013}.  \citet{Faerman2016} include clumping in
their models, but the higher density clumped gas is cooler, traced by
\ovi and the hotter gas, probed by \ovii and \oviii is indeed smooth.

To conclude, the hot gaseous CGM of the Milky Way appears to be diffuse
and extended. The CGM is clearly anisotropic as shown by the
distributions of both absorption and emission measurements. We have
determined the properties of the hot CGM by combining absorption and
emission measurements along three sightlines, two of which were
presented in Paper-II and one is presented in this paper. Additionally
in this paper we have fitted the absorption spectrum with a theoretical
collisional-ionization model, obtaining tight constraints on the
temperature of the gas. We have also added a $\beta-$model density
profile and show that the CGM remains diffuse, extended and massive. Our
results are consistent with numerical simulations as well as analytic
models and suggest that a large fraction of the MW missing baryons
reside in its hot CGM.

\section{Acknowledgments}

This work is supported in part by the NASA grant NNX16AF49G to SM. 
Y.K aknowledges support from grant DGAPA PAIIPIT IN104215 and 
CONACYT grant168519.

\clearpage

\begin{table}[tp]\footnotesize
\caption{Summary of observations for {\it Off-field2} and {\it Mrk~509} }
\begin{tabular}{lcccccc}
\hline
Experiment & Target  & OBSID  &  Start Date  & Exposure & $l$ & $b$ \\
           &         &        &              &   (ks)   &  {\it deg}  & {\it deg} \\
\hline
{\bf XMM-Newton}& {\it Off-field2}  & 722310201 & 11/19/2013 & 63 & 37.4  & -30.6 \\
{\bf Suzaku }   & {\it Off-field2}  & 509043010 & 05/07/2014 & 61 & 37.4 & -30.6 \\
{\bf Chandra}   & {\it Mrk~509}     & 13864     & 09/04/2012 & 170 & 35.9 & -29.8 \\
                & {\it Mrk~509}     & 13865     & 09/07/2012 & 99  & 35.9 & -29.8 \\
\hline
\end{tabular}
\end{table}

\clearpage

\begin{table}[tp]\footnotesize
\caption{$O_{VII}$ and $O_{VIII}$ emission lines intensities}
\begin{tabular}{lcc}
\hline
Dataset  &  $O_{VII}$    &  $O_{VIII}$   \\
         & $ph~s^{-1}~cm^{-2}~Sr^{-1}$ & $ph~s^{-1}~cm^{-2}~Sr^{-1}$\\
\hline
XMM-Newton & $19.8\pm0.4$ & $8.1\pm0.8$ \\
Suzaku  &  $30.9\pm6.6$  &  $7.8\pm1.5$ \\
\hline
\end{tabular}
\end{table}

\clearpage

\begin{deluxetable}{lcccc}
\tabletypesize{\scriptsize} \tablecaption{Z=0 Absorption lines observed in
the Mrk509 \chandra HETG-MEG spectra.}  
\tablewidth{0pt} \tablehead{
\colhead{$\lambda_{obs}$} & \colhead{EW} & \colhead{EW$^{~a}$}   & \colhead{Ion Name} & \colhead{$\lambda_{rest}$}\\
\colhead{$m\AA$}         & \colhead{m$\AA$}   & \colhead{m$\AA$}   &\colhead{} & \colhead{$\AA$} 
}  
\startdata 

$21.61\pm0.01$           &  $19.5\pm4.8$    &  $23.9\pm5.0$    &  \ovii $\alpha$ & 21.602  \\
$18.62\pm0.01$           &  $11.0\pm4.0$    &  $11.7\pm4.1$    &  \ovii $\beta$  & 18.627  \\
$18.97\pm0.01$           &  $12.0\pm3.4$    &  $10.3\pm4.3$    &  \oviii $\alpha$& 18.969  \\
\enddata
\tablenotetext{a}{Mrk509 Chandra LETG measurement} 
\end{deluxetable}

\clearpage

\begin{deluxetable}{lc}
\tabletypesize{\scriptsize} \tablecaption{Best fit PHASE parameters}
\tablewidth{0pt} \tablehead{
\colhead{Parameter} & \colhead{z=0 Component}}
\startdata 
$\log {\rm T (K)}$            & $6.33\pm0.16$  \\
$\log N_{H} (cm^{-2})$   & $20.33\pm0.19$$^{a}$ \\
$b~km~s^{-1}$           &   $74\pm80$    \\        
$N(O_{VII})~ cm^{-2}$    &  $(2.35\pm0.4)\times 10^{16}$  \\
$N(O_{VIII})~ cm^{-2}$   & $(1.81\pm0.5)\times 10^{16}$   \\
$N(Ne_{IX})~ cm^{-2}$    &  $(7.14\pm1.8)\times 10^{15}$  \\
\enddata
\tablenotetext{a}{assuming $Z=0.3Z_{\odot}$} 
\end{deluxetable}

\clearpage

\begin{deluxetable}{lc}
\tabletypesize{\scriptsize} \tablecaption{Baryonic Mass Estimates of the 
Milky Way}
\tablewidth{0pt} \tablehead{
\colhead{Component} & \colhead{Mass ($M_{\odot}$)}}
\startdata 
Virial Mass                &    $1 \times 10^{12} M_{\odot}$  \\
Baryonic Mass$^{a}$         &   $1.7 \times 10^{11} M_{\odot}$ \\
Stellar $+$ cold gas mass$^{b}$    &   $6 \times 10^{10} M_{\odot}$     \\
Missing Baryonic Mass      &   $1.1 \times 10^{11} M_{\odot}$ \\
{\bf Measured Hot halo mass}        &                              \\
Average; uniform density  (Paper I)  &   $> 6.1 \times 10^{10} M_{\odot}$  \\
This paper: uniform density     &  $3.9 \times 10^{10} M_{\odot}$ \\
This paper: $\beta$ model       &  $3.2-10 \times 10^{10} M_{\odot}$ \\
\enddata
\tablenotetext{a}{Calulated using cosmological baryon fraction of 
$f_{b}=0.17$ measured by the Wilkinson Microwave Anisotropy Probe 
\citep{Dunkley2009}}
\tablenotetext{b} {\citep{Sommer2006}}
\end{deluxetable}

\clearpage

\begin{figure}
\epsscale{.80}
\plotone{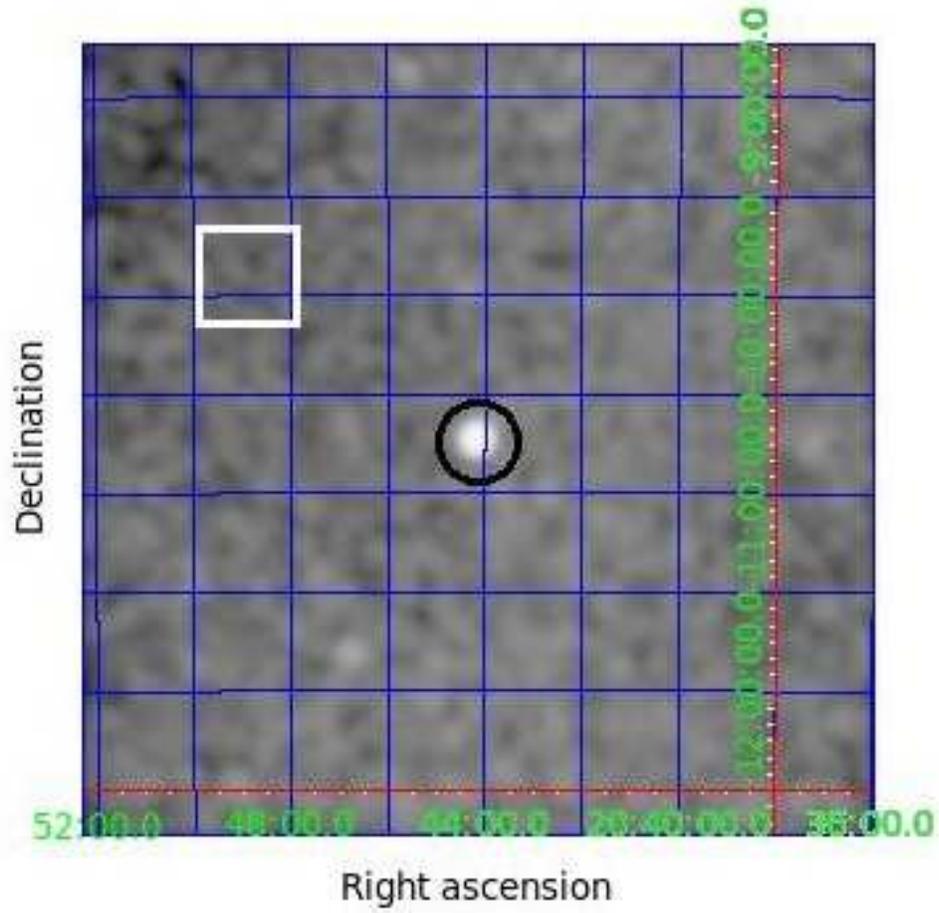}
\caption{RASS 3/4 keV band X-ray map in the vicinity of Mrk~509 
(black circle). The white square marks the nearby blank X-ray 
sky field ({\it Off-field2}).}
\label{fig1}
\end{figure}

\clearpage
\begin{figure}
\begin{center}
\includegraphics[width=10cm,height=10cm]{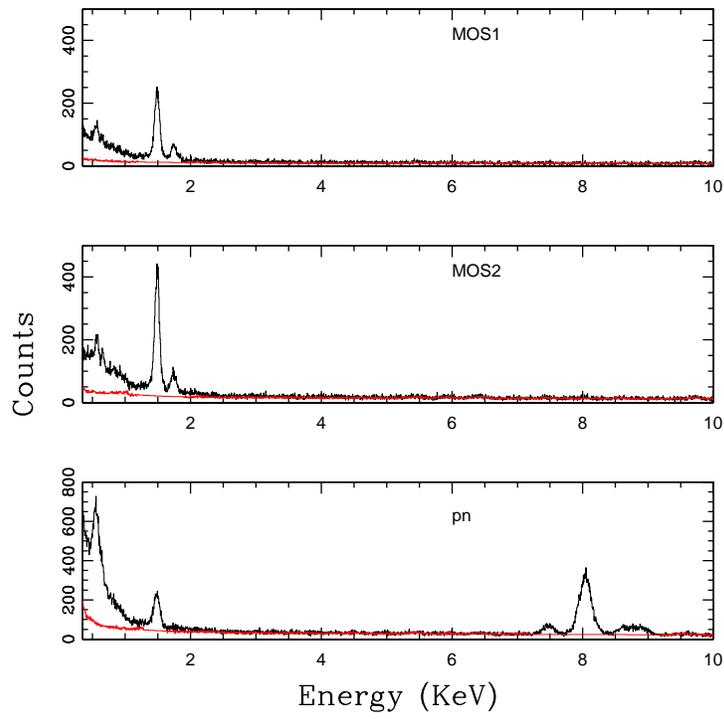}
\end{center}
\caption{XMM-Newton MOS and pn  spectra (black curves) of 
{\it off-filled2} extracted after removing the point sources. 
Red curves corresponds to background/QPB spectra.}
\label{fig3}
\end{figure}

\clearpage

\begin{figure}
\begin{center}
\includegraphics[width=10cm,height=10cm]{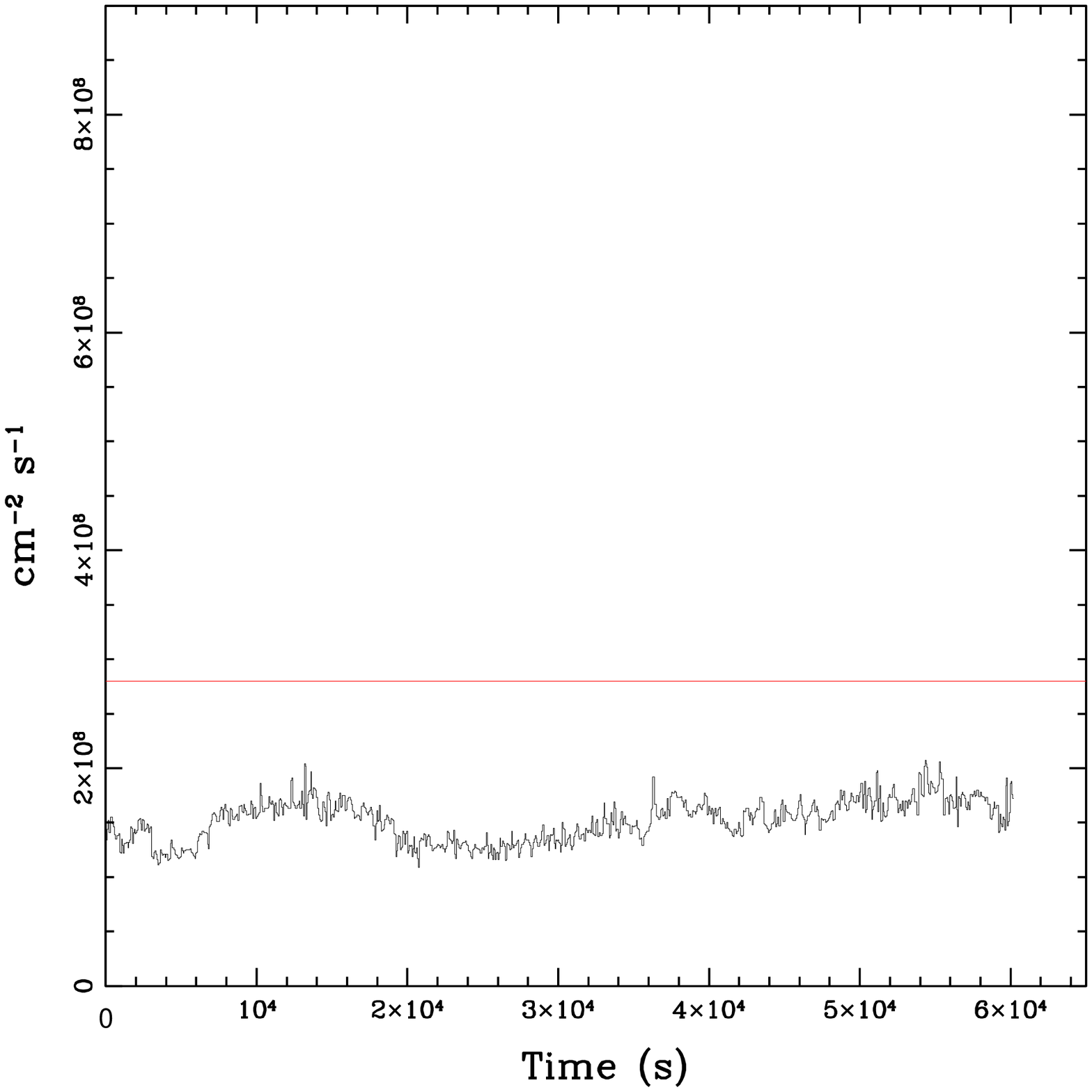}
\hspace{1cm}
\includegraphics[width=10cm,height=10cm]{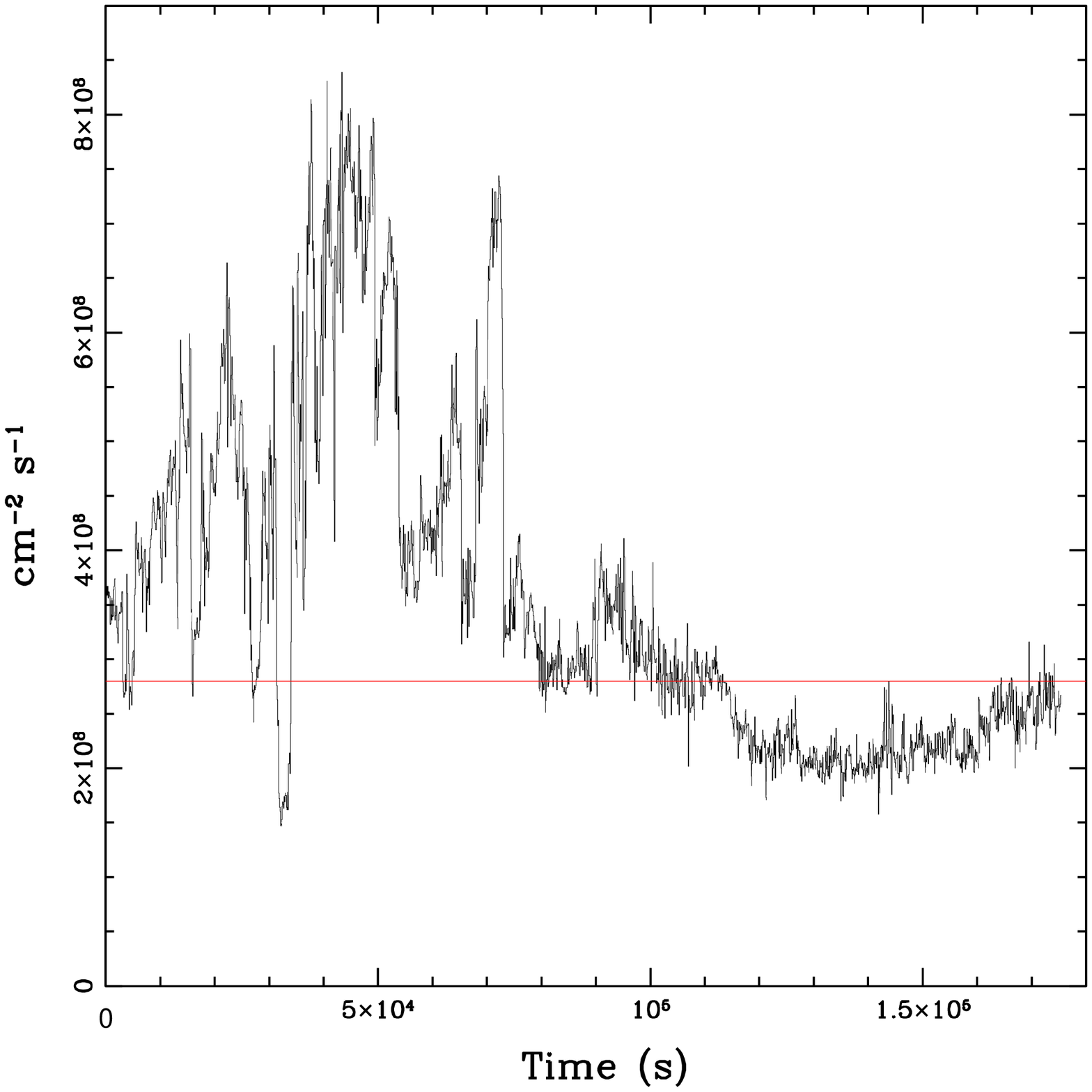}
\end{center}
\caption{Solar wind proton flux data from OMNIWeb during XMM-Newton (top) and 
Suzaku (bottom) observations of {\it Off-field2} periods. 
The {\it dashed} line shows the average proton flux at 1~AU 
($2.8 \times 10^{8}~$cm$^{-2}$~s$^{-1}$).}
\label{fig2}
\end{figure}

\clearpage

\begin{figure}
\begin{center}
\includegraphics[width=10cm,height=10cm]{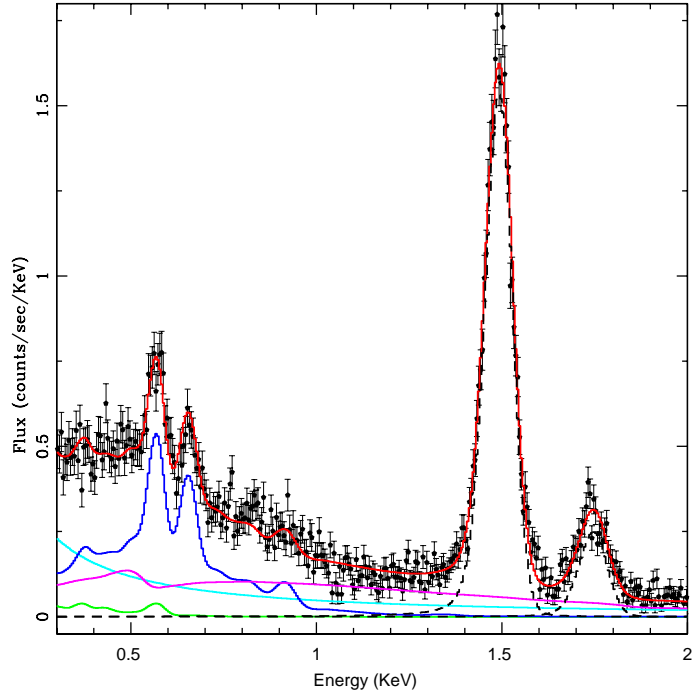}
\end{center}
\caption{MOS2 {\it Off-field2} spectrum, with the best-fit model
  ($\chi^{2}/d.o.f. = 1.1$, red curve).  We also plotted the individual
  components of SDXB: foreground component (LB+SWCX: green curve),
  extragalactic background (unresolved point sources: magenta curve) and
  galactic halo (blue curve). The cyan curve corresponds to residual
  partical background (modeled as unfolded power-law) and the black
  dashed gaussians model the instrumental Al and Si lines. }
\label{fig3}
\end{figure}

\clearpage

\begin{figure}
\epsscale{.80}
\plotone{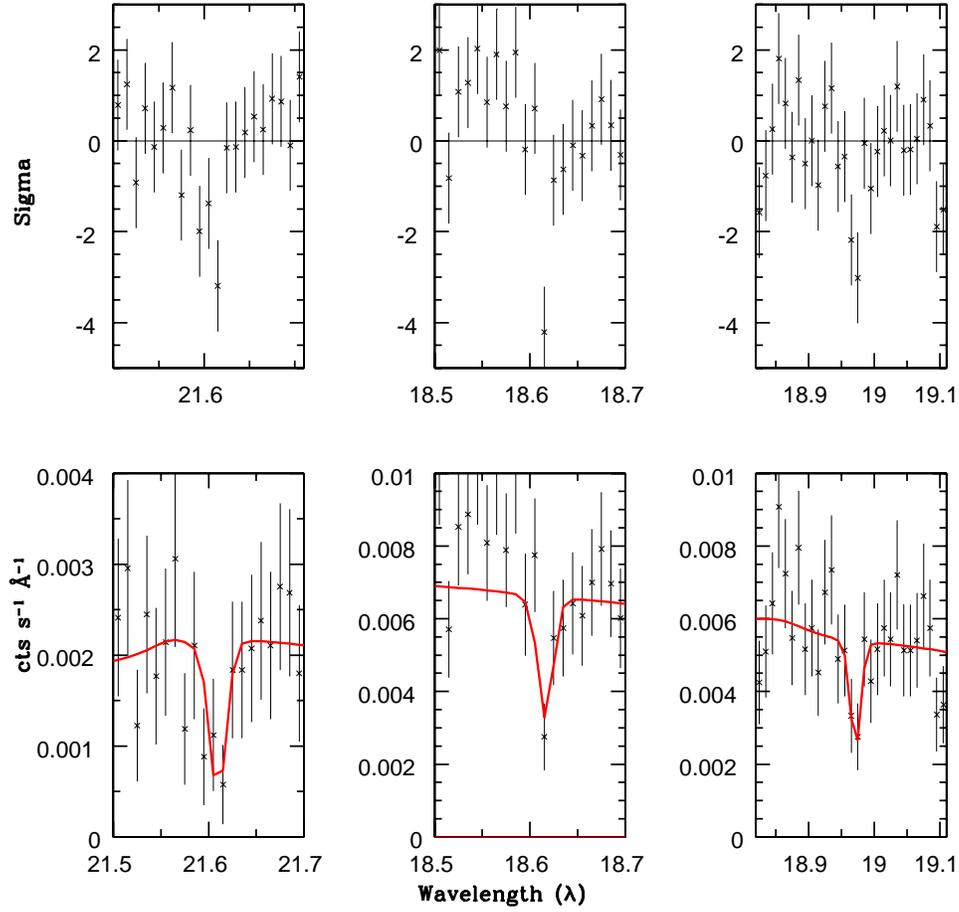}
\caption{Mrk509 Chandra HETG spectrum. \ovii K$\alpha$, K$\beta$, and
  \oviii K$\alpha$ absorption lines are clearly detected and are fitted
  with gaussian profiles (bottom panel). In the top panel residuals to the 
continuum model are shown.}
\label{fig4}
\end{figure}

\clearpage

\begin{figure}
\epsscale{.80}
\plotone{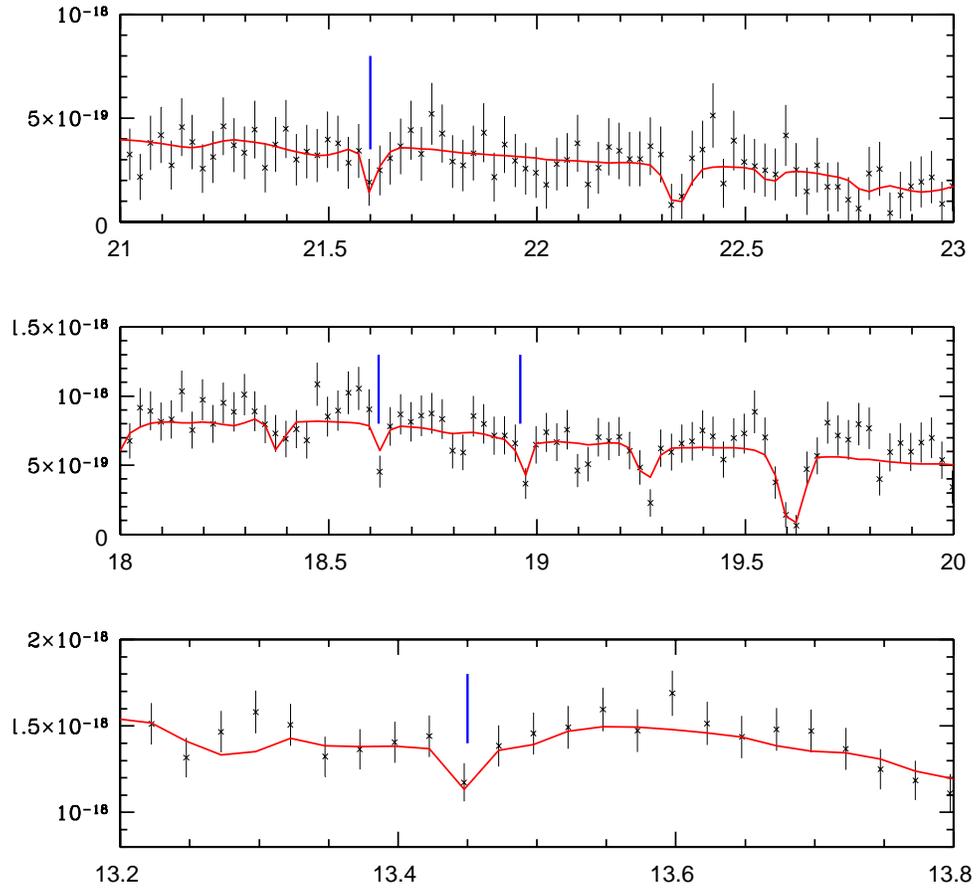}
\caption{PHASE fit to the HETG spectrum of Mrk509. $z=0$ absorption
  lines are marked; the rest are from the intrinsic warm absorber. }
\label{fig4}
\end{figure}

\clearpage

\begin{figure}
\epsscale{.80}
\plotone{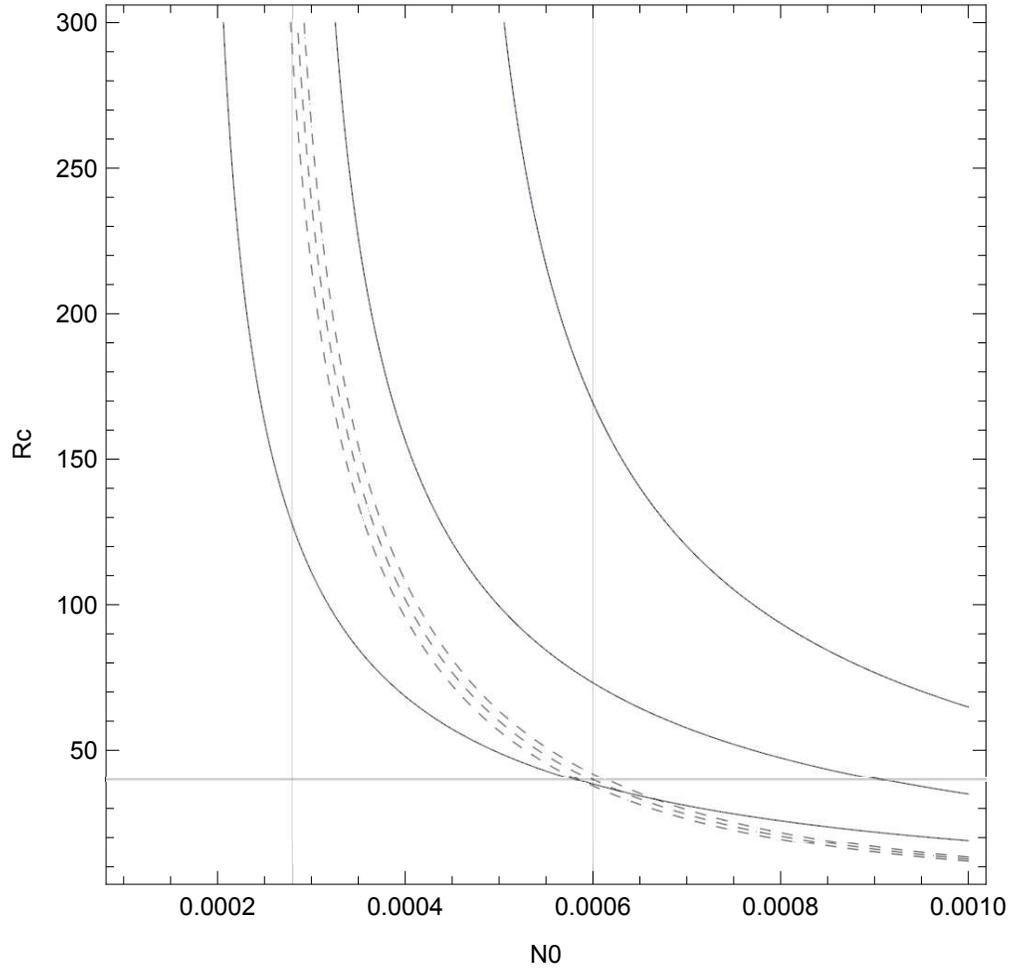}
\caption{A $\beta-$model density distribution fit to data. The 
central density N$0/\mu$ (in $cm^{-3}$) is plotted on X-axis while 
the Y-axis plots core radius R$_c$ (in Kpc). The solid lines match the
  best-fit and $1\sigma$ error values of the observed column
  density. The dashed lines are for the emission measure. Verticle lines show 
  the allowed range in N$_{o}$ and the horizontal line shows the 
  lower limit on the core radius.}
\label{fig4}
\end{figure}

\clearpage

\begin{figure}
\epsscale{.80}
\plotone{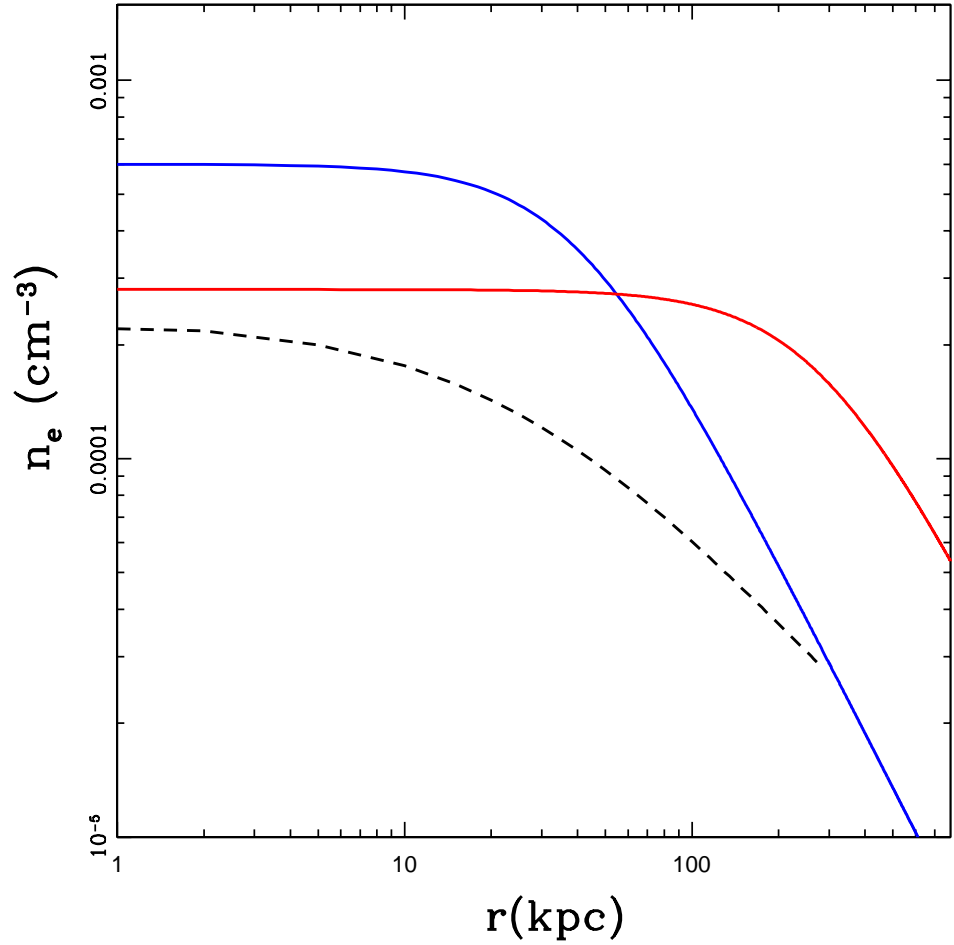}
\caption{The $\beta-$profiles deduced from this work are shown with
  solid lines. The blue and red curves correspond to the upper and lower
  limits on density, respectively. The dashed black curve is the Maller \&
  Bullock profile from Fang et al. (2013). }
\label{fig4}
\end{figure}

\end{document}